# Reduced graphene oxide for control of octahedral distortion in orthorhombic *Pbnm* perovskite

*Vidyarajan N.[a], L.K.Alexander[a,\*], Abraham Joseph[b]*

[a]Department of Physics, University of Calicut, Kerala, India-673635.

[b]Department of Chemistry, University of Calicut, Kerala, India-673635.

*Corresponding author:
Email address: LKA@uoc.ac.in (Libu K. Alexander)

*Abstract:*

Perovskites and related structures are renowned for technologically useful applications. The structural variations in perovskite crystals result in novel properties which can be regulated by external forces. The study attempts to regulate distortion of a perovskite compound $LaFeO_3$ driven solely by reduced graphene oxide (RGO) and to demonstrate the effect using a critical parameter - electrochemical conductivity. A series of graphene concentration varied nanocomposites were synthesized *via* hydrothermal method. The crystal distortion is investigated using X-ray diffraction and Rietveld refinement. By analysis of the oxygen and Fe ion peaks of X-ray photoelectron spectroscopy, along with Raman spectra, a clear understanding of the distortion process is developed. The remarkable effect of distortion on dispersed phases generated by graphene matrix is coined with a new phrase "*Graphenostortion*". The study revealed that the environment for the graphenostrotion originated from the interaction of oxygen functional groups of RGO with $LaFeO_3$. The variation in bond covalency influenced by the percentage concentration of non-lattice oxygen, distorted the *Pbnm* geometry of the graphene nanocomposite. The developed mechanism for distortion is in good agreement with electrochemical activity studies. RGO as a tool to control distortion in metal oxide could facilitate new facets in technological applications of graphene nanocomposites.

## Introduction

Perovskites, an adaptable crystalline framework are extremely attractive regarding the fundamental and technological aspects. The variety of properties in electric, magnetic and structural fields is based on the ability of the perovskites to accommodate a wide range of elements, oxidation states, and stoichiometries irrespective of the size of the cations. Properties can be modified by varying the concentration of constituent phases of the nanocomposite, which in turn gives an insight into the correlation between structure and the functional properties. The strong electron-lattice correlation in the perovskite metal oxides implies that structural distortion can lead to variation in physicochemical properties [1-3]. The distortion in the $BO_6$ octahedra, embedded within the crystal lattice, tuned by changing size (bond length (B-O)), shape (no. of unequal bonds) and connectivity can affect different properties of the perovskite compounds [4]. The chemical substitution within the lattice and the exertion of pressure during formation of the crystals are reported as leading strategies for changing the size and shape of the octahedra [5-9]. Epitaxial strain effect is reported to be one of the mechanisms to alter the connectivity between the octahedral frameworks [10-11]. The variation in three-dimensionally connected $BO_6$ octahedra by the aforementioned methods, in turn, leads to distortion in the global crystal structure of the material. Still, one of the attractive and challenging areas in engineering the functionality of a perovskite is to finely regulate the structural distortion without changing the constituents of the perovskite.

Graphene formed out of oxidation-exfoliation-reduction of graphite, create reaction sites like hydroxyl, carboxyl, carbonyl and epoxy groups, enabling a platform for engineering wide range of functionalities. On account of its outstanding properties, tailoring various nanostructures on graphene surface is of great interest and has applications in fields like chemical sensors, fuel cell, supercapacitors, batteries and photocatalysis[12-14].

There are reports showing the enhanced functionality of metal oxides driven by the combination with graphene[15-18]. Because of the attractive features such as catalytic behavior, sensing activity, thermoelectric properties and photovoltaic enhancement, nowadays serious attempts are made to engineer the electrical characteristics of graphene-perovskite hybrid materials [19-22]. The crystal parameters, including distortion, hold a significant role in deciding functionality of any material especially in perovskite [23,24]. But a detailed understanding of the features and the origin of structural variation in the dispersed phase impelled by graphene is to be developed.

Here, by utilising the surface properties, functionalisation of Reduced Graphene Oxide (RGO) with one of the significant member from orthoferrite perovskite family $LaFeO_3$ (LFO) has been carried out. Again, this work attempts to regulate the distortion of a perovskite compound propelled solely by using RGO and to demonstrate the effect using a critical parameter, electrochemical conductivity. The structural variation is established using Rietveld refinement and Raman spectroscopy. The distortion affects the oxygen concentration within the lattice, which is quantified using X-ray Photoelectron Spectroscopy (XPS). A fundamental understanding of the origin of the distortion accomplished in LFO is discussed.

## 2. Experimental Section

### 2.1. Sample preparation

GO was prepared by the modified Hummer's method [25] and LFO was formed by citric acid mediated solution method [26]. The Graphene–Lanthanum iron oxide (GLFO) nanocomposite was synthesised by one-pot hydrothermal process. The prepared GO was dispersed in water and subsequently, LFO was mixed with the GO solution in different weight ratios. The gram equivalent ratios of GO to LFO was 1:10, 5:10 and 10:10 and were named as GLFO-1, GLFO-5, and GLFO-10 respectively. The solution mixture was subjected to hydrothermal treatment at $180^0$ C for 6 hours. The hydrothermal process could lead to combined reduction and functionalisation of graphene oxide (GO). The black precipitate obtained was separated by centrifugation and dried at $60^0$ C.

### 2.2. Characterization

X-ray diffraction (XRD, Cu K Rigaku-600) and Raman spectroscopy (HORIBA-532nm laser) were used for the structural investigation. The morphological examination was done using field emission scanning electron microscopy (FESEM) (GEMINI-ZEISS). The electrochemical workstation (BIOLOGIC SP-150) was used for the study of electrochemical conductivity. For all electrochemical measurements, a three electrode configuration was used with Pt as the counter electrode and saturated calomel electrode as reference. The graphite working electrode was drop casted with 5μL (concentration: 2mg/ml) of graphene nanocomposite and dried at $80^0$C. The current variation driven by nanocomposite coating was investigated with ammonium (AM) ions using differential pulse voltammetry (DPV) studies. Using XPS, the concentration of element with different oxidation state were quantified in order to study the oxygen distribution in the nanocomposite.

## 3. Result and discussion

The XRD pattern of crystalline LFO as shown in Figure 1 was completely indexed in agreement to the orthorhombic *Pbnm* structure having highest intensity for (121) plane [JCPDS (01-078-4429)][27]. Figure 1 also shows the XRD patterns for graphene-LFO nanocomposites

(GLFO). In GLFO-10, the presence of an additional phase of $Fe_2O_3$ is identified. In order to understand the RGO concentration dependent dynamics of the crystalline structure, Rietveld refinements were carried out on LFO and GLFO compounds. The fits are shown in supplementary material Figure S1 to S4. Refinement on entire GLFO composite has been done based on *Pbnm* space group which has a lower symmetry than cubic *Pm3m*. The unit cell parameters obtained by refinement are shown in Table 1. The table suggests a significant dependence of the lattice parameter on the RGO concentration.

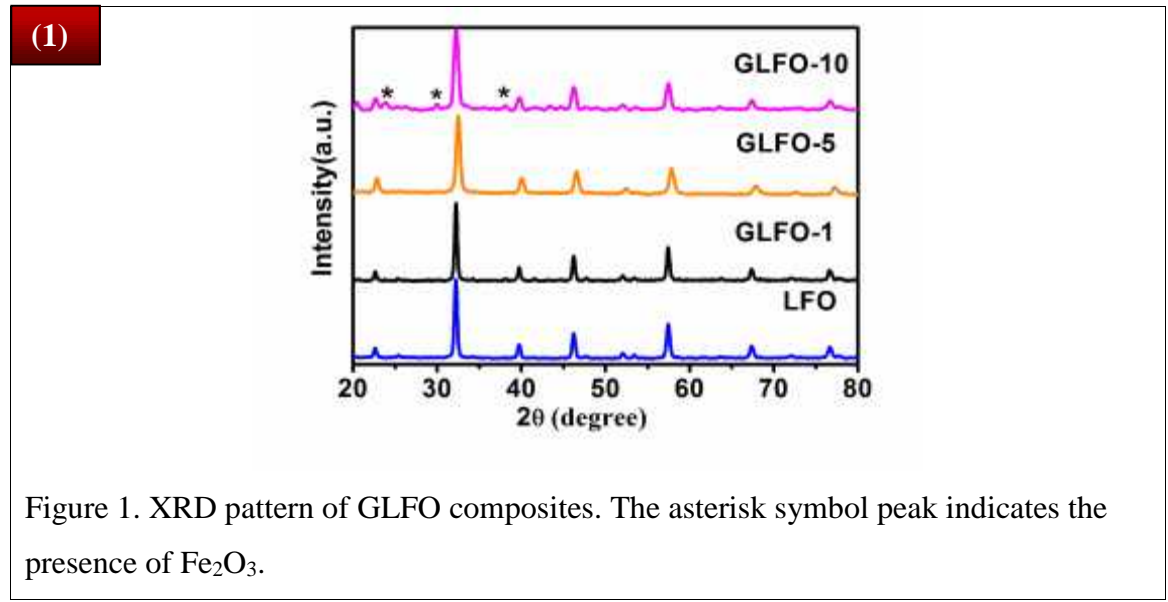

Figure 1. XRD pattern of GLFO composites. The asterisk symbol peak indicates the presence of $Fe_2O_3$.

Table 1. Rietveld refined parameters for LFO and GLFO composites (Based on supplementary material Figure S1 to S4).

| Parameters | LFO | GLFO-1 | GLFO-5 | GLFO-10 |
|---|---|---|---|---|
| a (A$^0$) | 5.567 | 5.574 | 5.554 | 5.559 |
| b (A$^0$) | 5.563 | 5.577 | 5.549 | 5.576 |
| c (A$^0$) | 7.868 | 7.873 | 7.851 | 7.892 |

Raman spectroscopy is an efficient tool for the investigation of structural distortion and oxygen motion in the crystal structure. The orthorhombic structure with a *Pbnm* space group has total 24 Raman active modes, which are $7A_g \oplus 7B_{1g} \oplus 5B_{2g} \oplus 5B_{3g}$[28]. The Raman peaks obtained from crystalline LFO are shown in Figure 2(a). The detailed Raman peak evaluation is given as supplementary information (S8). The main peaks of LFO at 431 cm$^{-1}$ and 626 cm$^{-1}$ originated from the stretching of Fe-O bond. The symmetry corresponding to these peaks are $A_g$ and $B_{1g}$ respectively [29, 30]. In GLFO-5, as shown in Figure 2(a), the peaks corresponding to $A_g$ got shifted to 442 cm$^{-1}$ and the $B_{1g}$ peak is broadened and shifted to 678 cm$^{-1}$. Again, as seen in Figure 2(b), the Fe-O bond vibration peaks show maximum shift for GLFO-5. Lattice strain due to expansion or compression of crystal structure leads to a shift in Raman peaks. Among GLFO composites, the maximum lattice compression is expected for GLFO-5 as suggested by the highest shift. The broadened peak at 678 cm$^{-1}$ shows splitting (as deconvoluted in supplementary Figure S5). The RGO concentration dependent changes in Raman peaks at 431 and 626 cm$^{-1}$ originates from dissimilar distortions in the crystal structure.

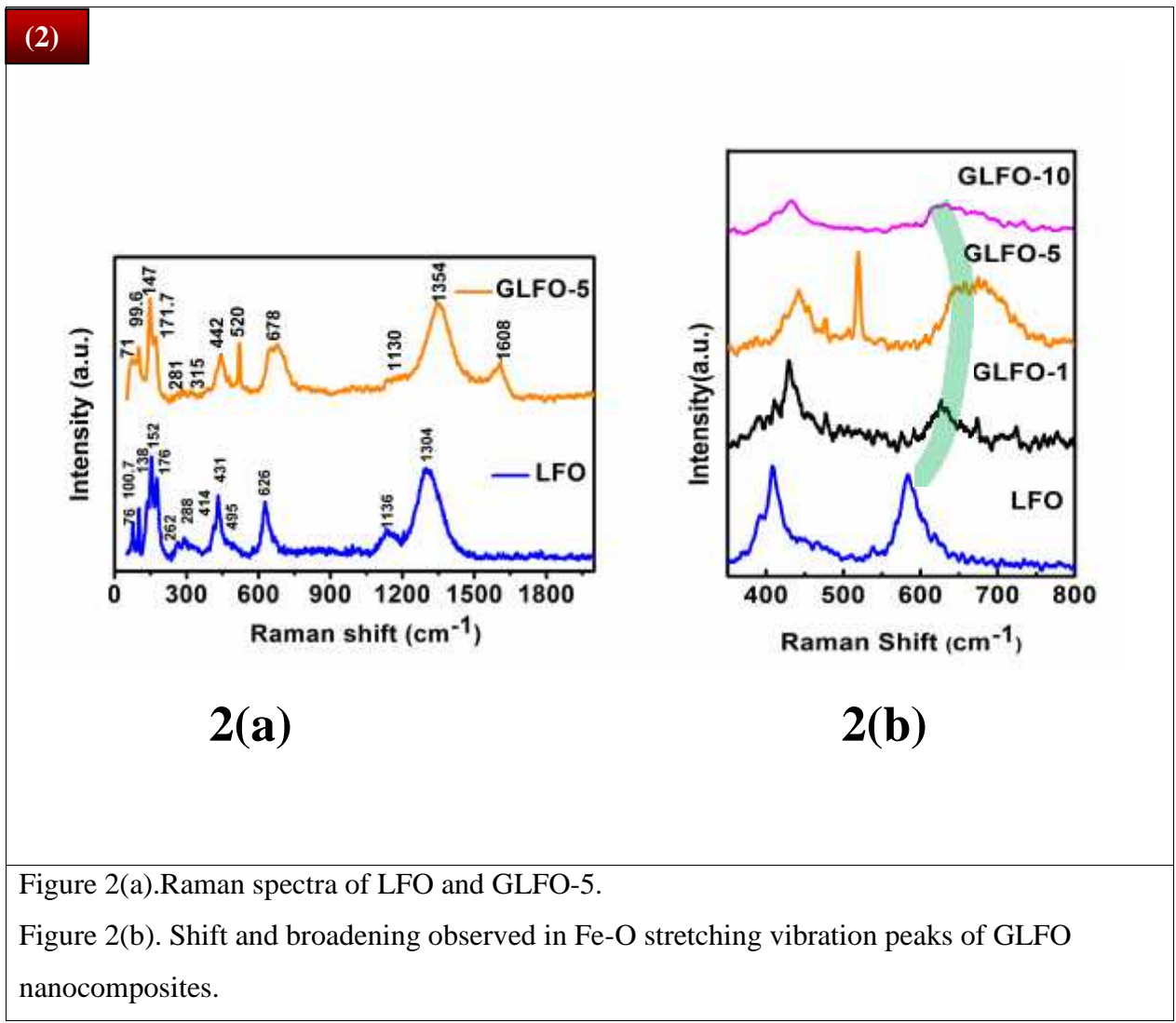

(2)

2(a)    2(b)

Figure 2(a). Raman spectra of LFO and GLFO-5.

Figure 2(b). Shift and broadening observed in Fe-O stretching vibration peaks of GLFO nanocomposites.

Thus, a graphene concentration dependent structural variation in GLFO composites is suggested by XRD pattern and confirmed by Raman spectroscopic analysis. Recently Yinlong Zhu *et al*. have studied the mechanism of tuning the cation deficiency in $LaFeO_3$ by utilising electrochemical investigations (OER and ORR) [31]. Also, Mefford *et al*. have explored the effect of bond covalency on doped $LaCoO_3$ by studying effective electrolysis of water [32]. Therefore, the electrochemical investigations can be hypothesised as a technique to demonstrate the curious structural variations in GLFO composites.

The electrochemical behavior of the RGO modified LFO was investigated by cyclic voltammetry (CV). The analysis was done in acetate buffer (0.1 M, pH 4.6) using three electrode assembly. The influence of ammonium (2.5nM AM) on the electrochemical behavior of modified electrodes (acetate buffer; scan rate: 50mV/s) is portrayed in Figure 4(a). As seen, a higher activity is shown by the RGO nanocomposite (GLFO) modified electrodes compared to LFO. This can be attributed to the higher surface area and defect density offered by RGO and the consequent increase in the electronic transfer between the analyte and the electrode.

Among the different modified electrodes, the peak current exhibited an increasing trend from LFO upto GLFO-5 and then got dropped down in GLFO-10. Thus GLFO-5 shows a peak electrochemical activity towards the detection of ammonium. Supplementary Figure S6 compares the CV of GLFO-5 with and without ammonium ions. The area under the curve shows an enhanced current density for GLFO-5 in presence of ammonium ions. The electrocatalytic action of GLFO-5 was further investigated at different scan rates in acetate buffer. As Figure 4(b) shows, the redox peak current due to the diffusion controlled processes increases as the scan rate increases.

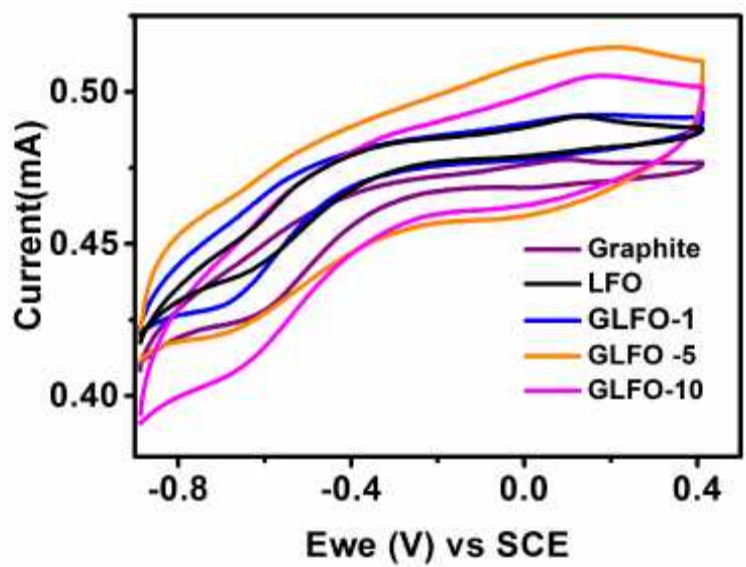

4(a)

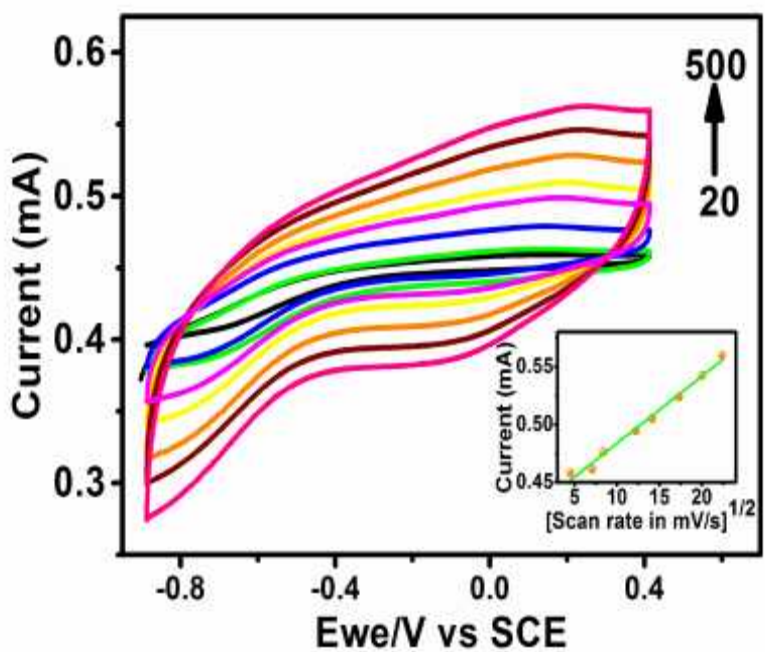

4(b)

Figure 4(a). Cyclic Voltammograms (CV) of Graphite, LFO, GLFO-1, GLFO-5 and GLFO-10 electrodes in 2.5nM AM.

Figure 4(b). CV of GLFO-5 at different scan rates. Inset: Calibration curve of GLFO-5

The variation in electrochemical conductivity of GLFO and the resultant parameters like the limit of detection and sensitivity of the electrodes were investigated using differential pulse voltammetry studies (DPV). Figure 5(a) shows the anodic peak current of various modified electrodes in presence of 2.5nM AM solution. Considering the enhanced DPV response shown by GLFO-5, further studies were performed on GLFO-5 electrode. The anodic peak current increases with the concentration of the analyte ranging from zero to 200 µM. The calibration curve (variation of current against concentration) is shown in the inset graph of Figure 5(b). An agreement with a single linear correlation of GLFO-5 modified electrode was found in the concentration range 0.27nM to 29nM; I (A) = (618 x Concentration in µM + 44.3x10$^{-6}$). The lower detection limit is found to be 0.636 nM and sensitivity is calculated as 2.12 nM from the graph.

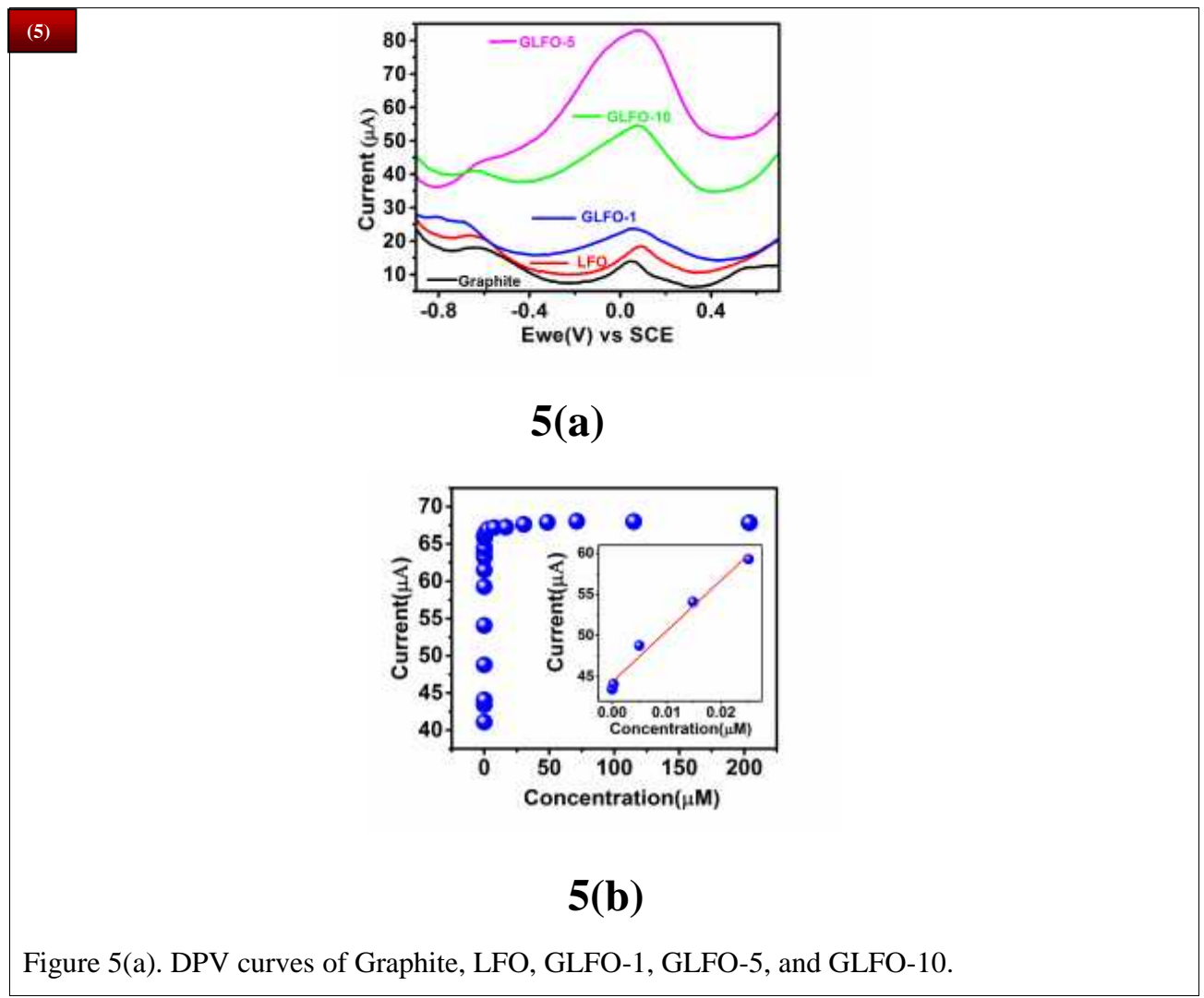

5(a)

5(b)

Figure 5(a). DPV curves of Graphite, LFO, GLFO-1, GLFO-5, and GLFO-10.

Figure 5(b). Variation of current with concentration for GLFO-5. The inset shows Calibration plot for the range 0.27nM to 30 nM. Fit equation: I (A) = 618 x Concentration in µM + 44.3x10$^{-6}$.

In order to understand the mechanism of enhanced sensitivity of GLFO-5 towards ammonium, the elemental analysis of the composite was carried out using X-ray photoelectron spectroscopy. The O1 XPS spectra in Figure 6 are asymmetric in nature and suggest the presence of more than two chemical states of oxygen binding energies. The peak at 528.9 eV in LFO is assigned to the lattice oxygen Fe-O and La-O in the perovskite structure as shown in Figure 6(a)[33]. But the lattice oxygen peak exhibited a shift up to 529.9 eV according to the concentration of RGO in the composite. The chemical shift indicates the influence of RGO on nature of metal-oxygen bond within the lattice. The evolutions of peaks corresponding to non-lattice oxygen components in GLFO were also analysed. In the graphene nanocomposite, peak at 531.08 eV is assigned to the oxygen doubly bonded to the aromatic carbon. The peak at 532.1 eV corresponds to the oxygen singly bonded with aliphatic carbon and the peak at 533.2 eV is assigned to oxygen singly bonded to aromatic carbon. The presence of intercalated adsorbed water molecule can be seen at 534.4 eV [34].

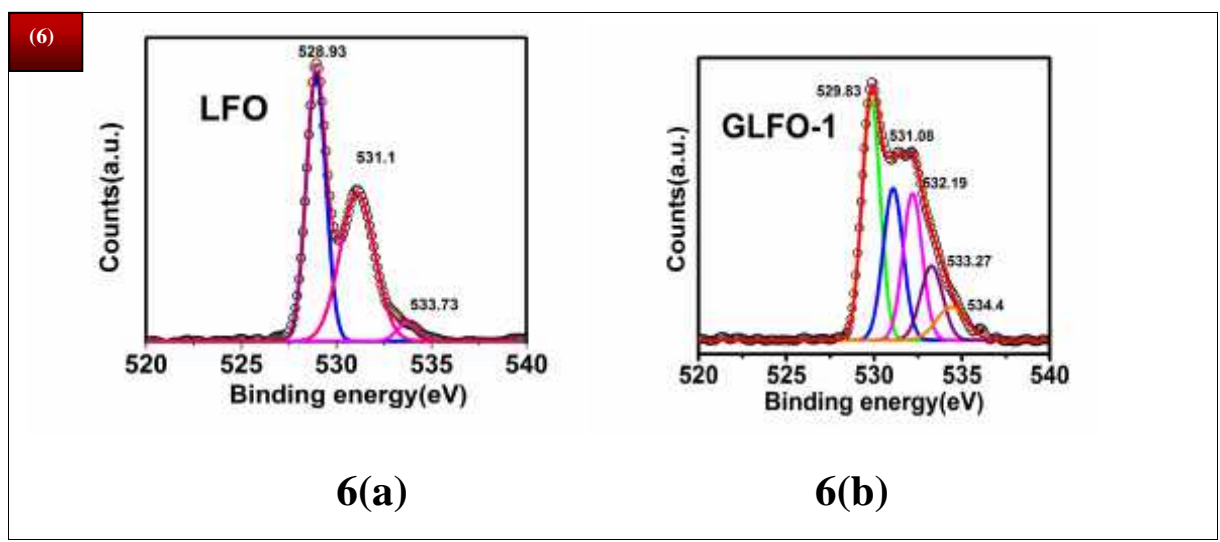

6(a)     6(b)

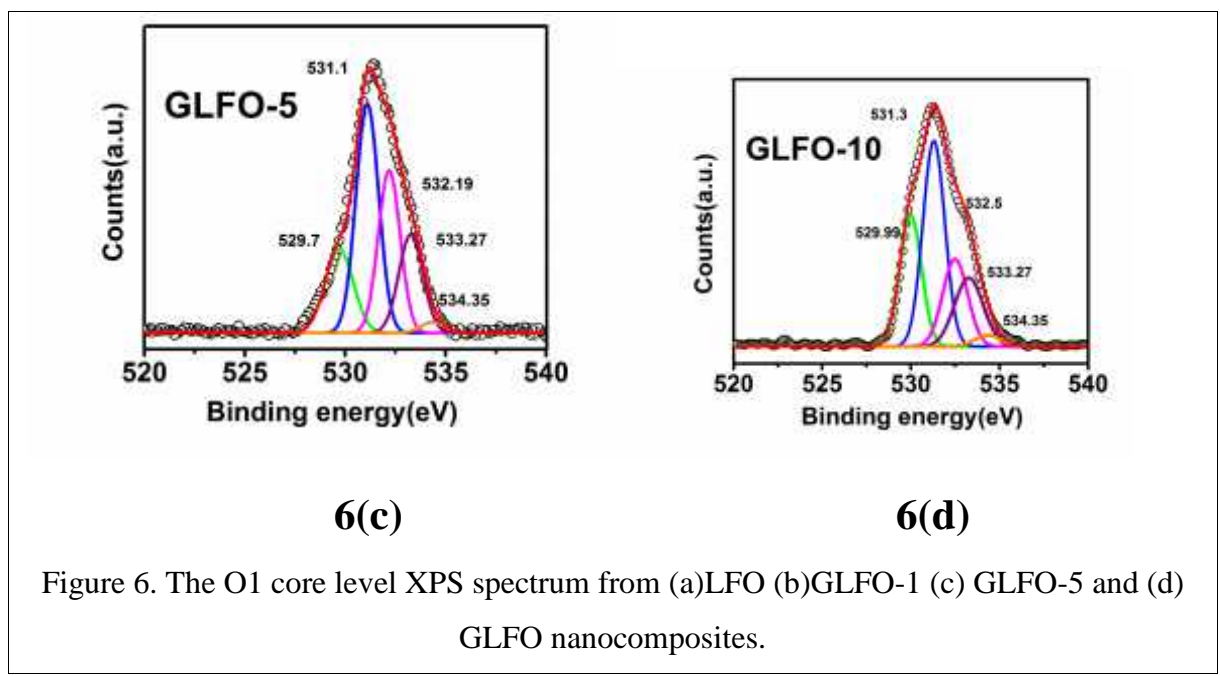

Figure 6. The O1 core level XPS spectrum from (a)LFO (b)GLFO-1 (c) GLFO-5 and (d) GLFO nanocomposites.

Therefore, GLFO nanocomposite shows a distinct RGO concentration dependence as observed in XRD and confirmed by Raman spectra results. Also, the dependence is clearly demonstrated by the electrochemical studies and XPS based elemental analysis on GLFO. A further task is to develop an understanding of underlying reasons for the highest dependence of RGO concentration on the intriguing behavior of GLFO-5.

The Figure 7 shows unit cell structure of LFO and various GLFO compounds plotted based on refined values from the Rietveld analysis. The analysis indicates that the incorporation of RGO played a critical role in tuning the extent of change in size, shape and co-operative tilting of octahedra. This change, in turn, can affect the electronic environment of ferrite metal ions in LFO.

*Change in Fe-O bond length and Fe-O-Fe bond angle*: The change in size and shape of octahedra was studied based on the change in Fe-O bond length and bond angle. The notations of different Fe-O bonds are shown in supplementary Figure S7. The effect of RGO on the bond length of Fe-O as shown in Table 2, indicates a clear dependence of the size and shape on RGO concentration. The $O^{2-}$ ligands attached with the Fe metal ion in LFO offer an electronic cloud distribution. In the case of LFO in GLFO composites, a cumulative effect of (1) electron rich functional groups in RGO and (2) the electron cloud from $sp^2$ bonded carbon in RGO produces an additional electric field over the Fe metal ions. The change in bond length modifies the overlap between the $3d$ orbitals of ferrite ion and $2p$ orbitals of oxygen ions, leading to change in electronic conductivity. Other than shape changes of the octahedra, Figure 7 emphasizes an inter-octahedral tilt and rotation through Fe-O-Fe in GLFO compared to LFO. The angle $\angle_{B-O-B}$ measures the tilting of the octahedron which is directly related to the cell distortion in the crystal structure. Ferrite ion is attached to six other octahedra through the Fe-O bond. So a small tilt in one $FeO_6$ octahedron leads to a distortion in the whole crystal structure of the perovskite material. Other than bond length, the tilting also affects the orbital overlapping. The Table 2 consolidate the overall variation of $\angle_{Fe-O-Fe}$ with the concentration of RGO. It indicates that GLFO-5 has the maximum deviation (decrease) in Fe-O-Fe angle and has a maximum rotation with respect to LFO. Figure 7(e) shows the individual O-Fe-O angles in single octahedron of GLFO-5 and LFO. It clearly demonstrates the enhanced distortion even in an individual octahedron of GLFO-5 in comparison to that in LFO. This enhancement is entirely contributed by graphene.

Table 2. Structural parameters of octahedron formed in LFO and GLFO composites

|  | Nanocomposite | | | |
|---|---|---|---|---|
|  | LFO | GLFO-1 | GLFO-5 | GLFO-10 |
| Bond length: | | | | |
| Fe-O1(x2) | 2.01 | 1.99 | 1.98 | 2.35 |
| Fe-O21(x2) | 2.02 | 2.05 | 1.97 | 2.19 |
| Fe-O22(x2) | 1.99 | 2.02 | 2.27 | 1.75 |
| Bond angle: | | | | |
| Fe-O-Fe | 157.6 | 150.3 | 135.6 | 170.1 |

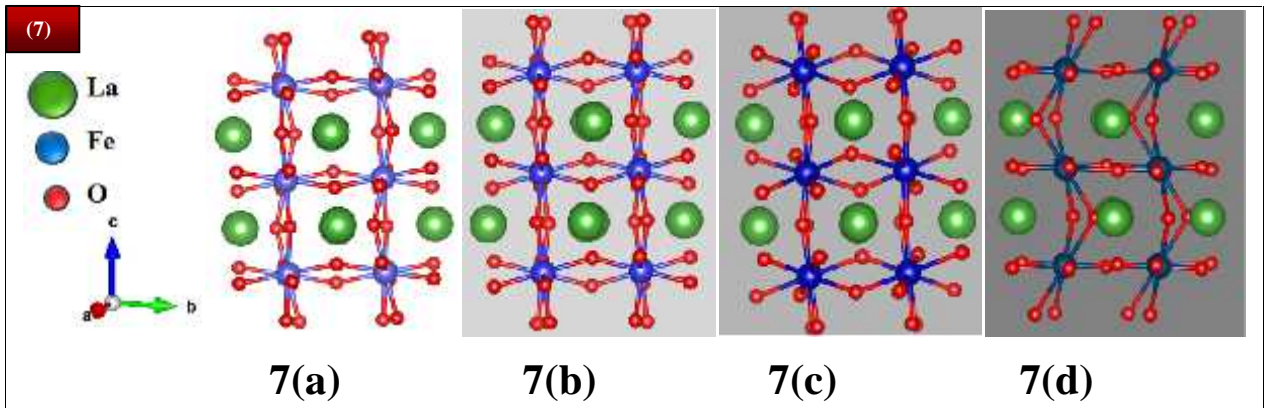

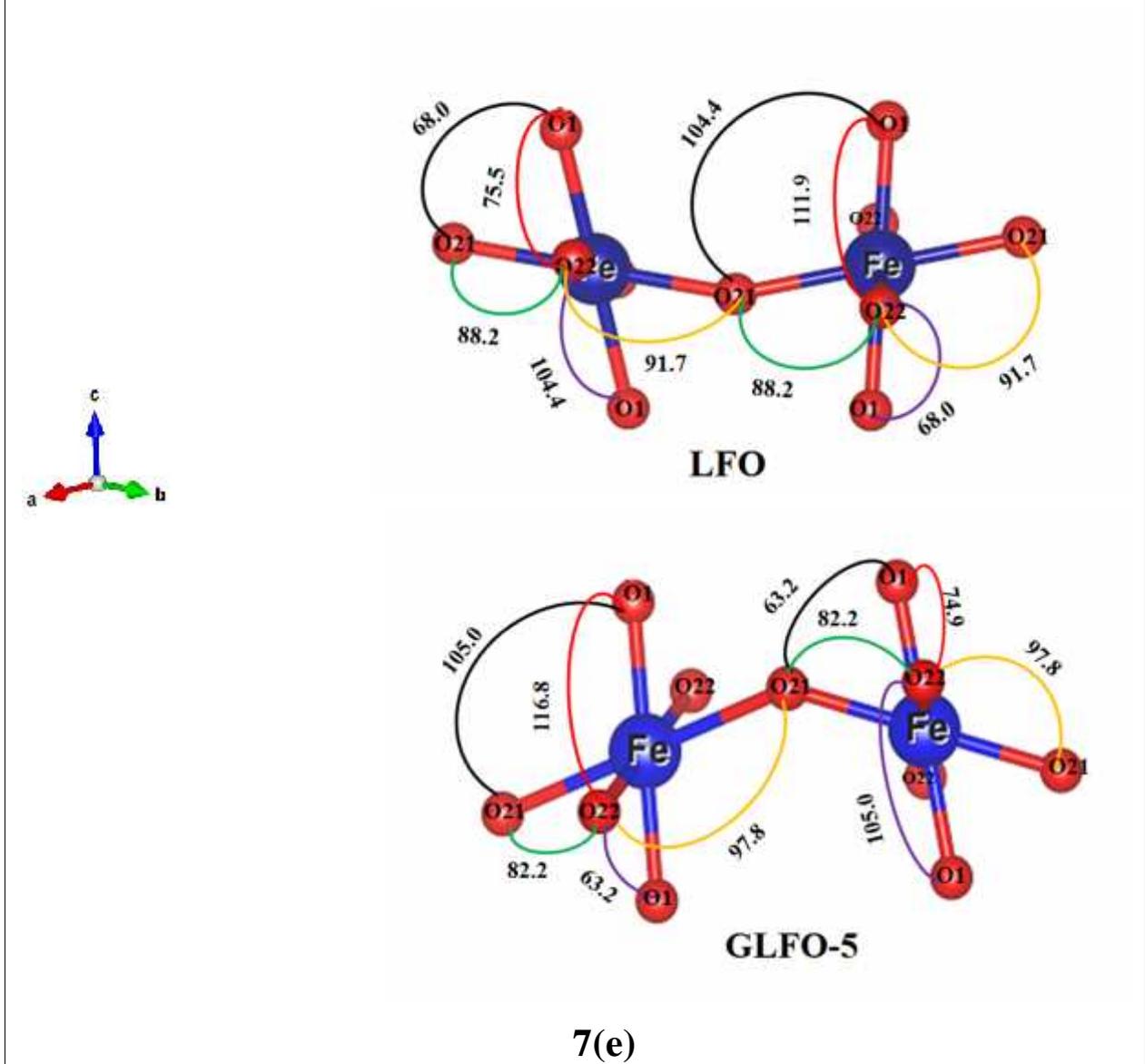

Figure 7. Structure of (a) LFO (b) GLFO-1 (c) GLFO-5 (d) GLFO-10. (The darkness of the background symbolizes relative concentration of RGO) (e) O-Fe-O bond angles of individual octahedron in LFO and GLFO-5

The noted change in bond length and angle can together be categorised as a distortion of the LFO perovskite. This distortion is clearly dependent on the graphene concentration and, therefore, this remarkable effect can be coined with a new phrase ***"Graphenostortion"***.

***Change in bond covalency:*** The electronic interactions play an important role in the distortion of a crystal lattice [35].Based on the pseudo-Jahn Teller effect an octahedral tilting could be implicitly generated by the change in bond covalency of the cation-anion bonds. Cammarata A. and Rondinelli J.M. report that the rotational angle and bond length imparts perturbations to the bond covalency of the composites [36]. Thus, the change in bond angle and bond length leading to an octahedral tilting simultaneously can lead to a change in bond covalency. Again, the bond covalency of the metal-oxygen bond is directly related to the bond valency [37].The bond valency obtained for LFO, GLFO-1, GLFO-5 and GLFO-10 using SPuDS calculation is 3.12, 2.98, 2.70 and 3.08 respectively [38]. As noted, the GLFO-5 has the lowest bond valency among GLFO composites and therefore the exhibited highest conductivity is expected. On further analysis specifically on bond valencies of ferrite ion of different Fe-O bonds, it was found that within GLFO-5, the Fe-O(2)1 bond could offer the best conductivity path. Clearly, in GLFO, the charged atmosphere offered by RGO led to change in covalency of the dispersed phase.

***Verification of distortion by Raman Spectra:*** The frequency shift observed in Raman spectra (Figure 2(b)) can be ascribed to the fact that the structural distortion affects the phonon-phonon interaction. The broadening of $B_{2g}$ mode at region 680 $cm^{-1}$ (Figure 2(b) and supplementary Figure S5) could indicate the presence of different oxygen species in the lattice. This can induce a change in polaraisability of oxygen ions leading to deviation in bond covalency.

***Distortion and non-lattice oxygen:*** Based on XPS, the effect of concentration of RGO is reflected in the concentration of lattice and non-lattice oxygen (Table 3). The analysis of Fe $2p$ peaks gives an idea that the interaction of reactive sites of GO with LFO induces a conversion of $Fe^{2+}$ into $Fe^{3+}$ions which is consolidated in Table 3. The oxygen species from RGO attack the oxygen deficiency created by the $Fe^{2+}$ ions within the crystal lattice. Mutually complementary effect of conversion of metal cation oxidation state and occupancy of non-lattice oxygen in the crystal influences the bond covalency of the metal-anion bonds. This affects a change in the size and shape of an octahedron. As noted in Table 3, the ratio of non-lattice oxygen to the lattice oxygen shows an increase with RGO concentration. This increase can be attributed to the oxygen offered by the functional groups of RGO. Beyond GLFO-5, the perovskite lattice can be assumed to be reluctant to accommodate more $Fe^{3+}$ ions in the lattice. Indicating the effect, the XRD spectrum shows the presence of $Fe_2O_3$ in the GLFO-10 composite. Simultaneously one can assume the presence of new oxygen species interacting with the dispersed phase of GLFO-5 as indicated by broadening in the Raman spectrum (Figure 2(b)&S5). The new oxygen species can be identified with the non-lattice oxygen. As also indicated by Raman spectra metal–lattice oxygen (Fe-$O_{lattice}$) bond covalency must have been influenced by the occurrence of non-lattice oxygen. Thus, RGO concentration dictated the percentage concentration of non-lattice oxygen.

Table 3. RGO dependent elemental concentration in LFO and GLFO nanocomposite

| Sample | Percentage concentration | | | |
| --- | --- | --- | --- | --- |
| | $Fe^{2+}$ | $Fe^{3+}$ | lattice oxygen | non-lattice oxygen |
| LFO | 36 | 64 | 49 | 51 |
| GLFO-1 | 22 | 78 | 34 | 66 |
| GLFO-5 | 17 | 83 | 17 | 83 |
| GLFO-10 | 3 | 97 | 24 | 76 |

Schematic representation of possible mechanism for the distortion in LFO driven by graphene or "Graphenostortion" is shown in Figure 8.

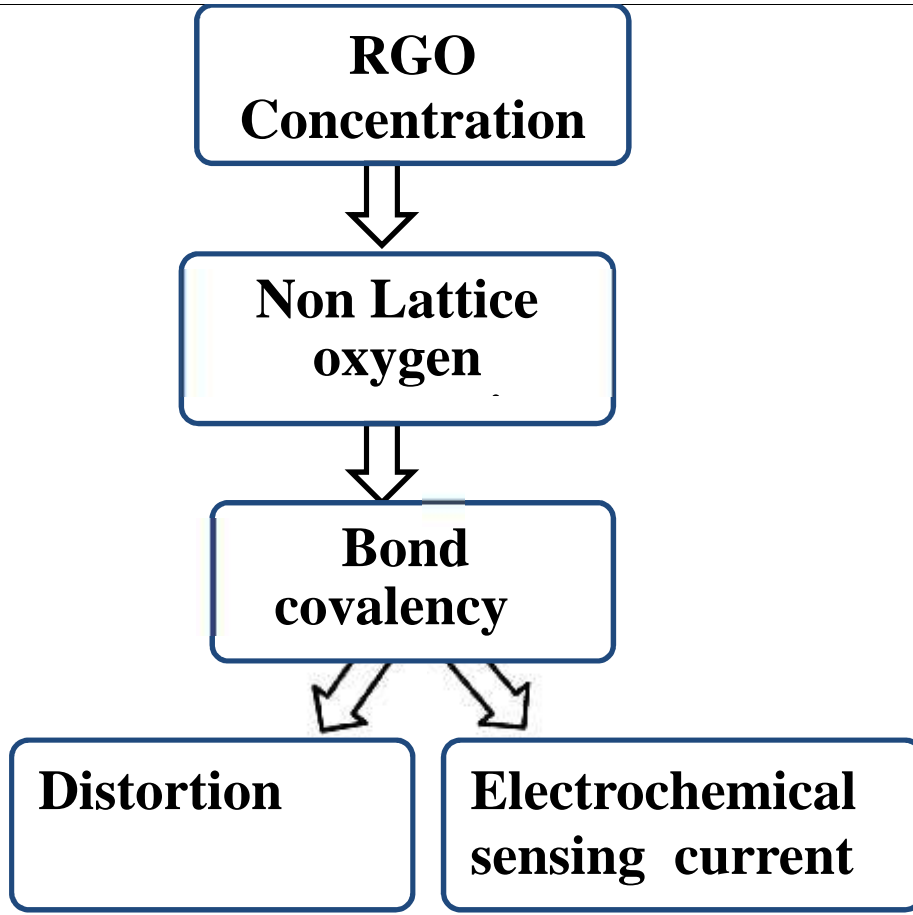

Figure 8: Schematic representation of a possible mechanism for the distortion in LFO driven by graphene ("Graphenostortion").

*Effect of distortion on electrochemical conductance:* Commenting specifically on ammonium sensitivity, the interaction of ammonium ions with the non-lattice oxygen within the composite determines the electrochemical activity. The interaction is in agreement with the following empirical formula of the reaction.

$$NH_4^+ + H_2O \rightleftharpoons NH_3 + H_3O^+$$

$$2NH_3 + 3O_{nl}^- \rightarrow N_2 + 3H_2O + 3e^-$$

The XPS analysis paved the way to find a connection between the highest current sensitivity in GLFO-5, oxygen concentration and the distortion impelled by RGO. GLFO-5 has the highest amount of non-lattice oxygen (Table 3). Remarkably, the Figure 9 suggests proportionality between the concentration of non-lattice oxygen and the sensing current. Therefore, the electrochemical conductance can be assumed to be fundamentally driven by the quantity of non-lattice oxygen in the composite.

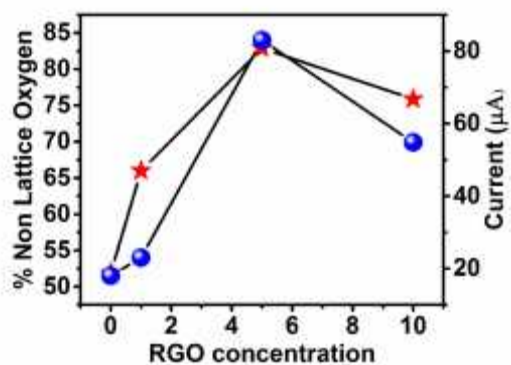

Figure 9. Variation of sensing current (dot) and percentage of non-lattice oxygen (star) with a concentration of RGO.

The present work indicates a possible mechanism of RGO interaction on orthorhombic crystals. The change in the physicochemical properties of the graphene-LFO composite can be attributed to the structural features of LFO, distribution of LFO on graphene sheets and the interfacial bonding between RGO and LFO. The synergistic effect between graphene and LFO in GLFO-5 among other RGO varied nanocomposites are explained on the basis of conversion of $Fe^{2+}$ to $Fe^{3+}$ ion created by the oxygen species. Thus, the modified bond covalency emerged due to the conversion of metal cation oxidation state, induced by the percentage of non-lattice concentration, has a key role in determining the distortion and regulation of electrochemical conductivity of the nanocomposite.

## 4. Conclusion

In summary, the distortion in perovskite $LaFeO_3$ is regulated by dispersing it over RGO matrix. Graphene induced structural distortion (Graphenostortion) on perovskite crystal structure is studied in detail. A series of graphene concentration varied nanocomposites were synthesized *via* hydrothermal method. The crystal distortion was indicated by XRD spectra and was verified by Raman spectroscopic studies. The structural deviations were quantified with the help of Rietveld refinement analysis and it indicates the nanocomposite with 5:10 ratio of GO to LFO shows higher deviation compared to LFO. The effect of distortion was demonstrated using electrochemical sensing studies on ammonium. By analysis of the oxygen and Fe ion peaks of XPS, along with Raman spectra, a clear understanding of the concentration of GO dependent distortion process was developed. The environment for the graphenostrotion originated from the interaction of oxygen functional groups of RGO with LFO. The non-lattice oxygen influence a change in concentration of $Fe^{2+}$ and $Fe^{3+}$ ions within the GLFO composite. Thus, the variation in bond covalency influenced by the percentage concentration of non-lattice oxygen in the composite creates inter octahedral and intra octahedron distortion and dictates the electrochemical conductivity. Graphenostortion opens up a new pathway for understanding the dynamics of metal cations and oxygen anions in perovskite. In addition, RGO as a tool to control the distortion in metal oxide could facilitate new spheres in technological applications of graphene.